\documentclass[10pt, conference, letterpaper]{IEEEtran}

\usepackage{cite}
\usepackage{amsmath,amssymb,amsfonts}
\usepackage[noend]{algorithmic}
\usepackage[pdftex]{graphicx}
\usepackage{textcomp}
\usepackage{xcolor}

\usepackage[hyphens]{url}
\usepackage{helvet}

\usepackage{algorithm}
\usepackage[caption=false]{subfig}

\usepackage{enumerate}

\usepackage{easyReview} 
\usepackage{balance}

\usepackage{amsthm}
\newtheorem{definition}{Definition}
\newtheorem{lemma}{Lemma}

\newtheorem{theorem}{Theorem}

\DeclareMathOperator*{\argmin}{arg\,min}

\begin{document}
\title{Dynamic Regret of Randomized Online Service Caching in Edge Computing}
\author{\IEEEauthorblockN{Siqi Fan, I-Hong Hou}
\IEEEauthorblockA{
\textit{Texas A\&M University}\\
College Station, USA \\
\{siqifan, ihou\}@tamu.edu}
\and
\IEEEauthorblockN{Van Sy Mai}
\IEEEauthorblockA{\textit{National Institute of Standards and Technology} \\
Gaithersburg, USA \\
vansy.mai@nist.gov}
}

\maketitle


\begin{abstract}
This paper studies an online service caching problem, where an edge server, equipped with a prediction window of future service request arrivals, needs to decide which services to host locally subject to limited storage capacity. The edge server aims to minimize the sum of a request forwarding cost (i.e., the cost of forwarding requests to remote data centers to process) and a service instantiating cost (i.e., that of retrieving and setting up a service). Considering request patterns are usually non-stationary in practice, the performance of the edge server is measured by dynamic regret, which compares the total cost with that of the dynamic optimal offline solution. To solve the problem, we propose a randomized online algorithm with low complexity and theoretically derive an upper bound on its expected dynamic regret. Simulation results show that our algorithm significantly outperforms other state-of-the-art policies in terms of the runtime and expected total cost. 
\end{abstract}



\section{Introduction}
\label{section:introduction}

Edge computing is a paradigm shift from cloud computing, where computation and data storage are brought closer to end users instead of offloading to a central cloud. This is done through the deployment of edge servers that can host (or cache) some popular services and process the corresponding computation tasks directly without having to forward them to remote clouds. Such close proximity provided by edge computing not only reduces bandwidth consumption in backhaul links, but also is critical for supporting various services and applications that require real-time data processing, such as augmented reality, virtual reality, and autonomous vehicles.

To fully realize the potential of edge computing in practice, several challenges in designing efficient service caching algorithms running on edge servers must be dealt with. First, edge servers can often host only a small number of services due to their limited storage capacity. Second, user requests are typically time-varying, and it is usually infeasible to fully predict future requests. Third, reconfiguring edge servers, which involves downloading necessary data and setting up virtual machines or containers, can incur significant delay and communication cost.

Existing studies for addressing these challenges typically design online policies that aim at learning and adopting an optimal static offline policy, e.g., Paschos \textit{et al.} \cite{paschos2019learning} and Zhang \textit{et al.} \cite{zhang2017proactive}. Here, a static offline policy is one that knows all future requests but only caches the same set of services at all times, and the cost difference between an online policy and the optimal offline counterpart is known as \textit{static regret}. Clearly, by focusing on learning the optimal static offline policy, these studies ignore potential gains from dynamically reconfiguring edge servers in response to changes in request arrival patterns. As a result, static regret is deemed less applicable when the environment is constantly changing. This motivates the notion of \textit{dynamic regret}, where an online algorithm is compared against optimal dynamic solutions in hindsight. Few recent studies \cite{chen2017online, li2020online} investigate dynamic regret for different applications but only design online algorithms that produce fractional solutions. Since service caching decisions are required to be integers, these algorithm cannot be applied directly. 

In this paper, we propose an online service caching policy with provably low dynamic regret by combining the strengths of two recently proposed algorithms, one is an online gradient algorithm \cite{li2020online} that has low dynamic regret but only produces fractional solutions and the other is a randomized algorithm \cite{fan2021online} that turns fractional solutions into integer ones but has no bounds on dynamic regret. We point out that this combination is not trivial because simply applying these two algorithms to our cost function does not readily lead to low dynamic regret due to the accumulated error from the randomization step. Thus, in order to bridge the gap between these two algorithms, we carefully construct an auxiliary function that not only admits fractional solutions but also explicitly incorporates the additional costs due to the randomized algorithm. Specifically, in each time slot, our algorithm first applies a projected gradient descent method to the auxiliary cost function using a customized efficient projection step. The output of this step is then treated as the probabilities of caching services at the edge server. Finally, a randomized algorithm is used to determine actual integer caching decisions. We also note that both algorithms in \cite{li2020online} and \cite{fan2021online} do not provide low complexity implementations of their projected gradient steps.

Our contributions in this paper are as follows. First, we develop an online service caching algorithm that yields integer solutions with provably low dynamic regret. In particular, we establish an upper bound of the regret that is sublinear in time when the path length, a measure of how frequently request arrival patterns change, is also sublinear in time. We prove that this upper bound can be further reduced when a finite window of request arrival predictions is available to the edge server. In addition, we develop a new algorithm for computing \textit{exact} projection onto a bounded simplex in nearly linear time; existing methods either run in quadratic time or only compute an approximate. This projection algorithm not only leads to an efficient implementation of our online caching algorithm, but is also of independent interest in other applications. Finally, simulation results show that our policy outperforms other state-of-the-art online algorithms under a variety of settings.

The rest of the paper is organized as follows. Section~\ref{section:related} reviews closely related work. Section~\ref{section:model} introduces our system model and the online caching problem of interest. Section~\ref{sec:alg} provides details of our randomized online service caching algorithm. Section~\ref{sec:DR} analyzes the expected dynamic regret of the algorithm. Section~\ref{sec:Com} proposes an efficient projection algorithm and analyzes the complexity of our randomized online algorithm. Some simulation results are given in Section~\ref{sec:simulation}. Finally, Section~\ref{sec:conclusion} concludes the paper.

\section{Related Work}
\label{section:related}

The majority of studies on the online caching problem are focused on static regret, which is evaluated by comparing with a static offline policy. For example, Paschos \textit{et al.} \cite{paschos2019learning}, Zhang \textit{et al.} \cite{zhang2017proactive}, Salem \textit{et al.} \cite{salem2021no} and Tan \textit{et al.} \cite{tan2015adaptive} form caching problems into online convex optimization and apply gradient method to obtain algorithms with sublinear static regret. 
Fan \textit{et al.} \cite{fan2021online} consider the problem of jointly optimizing service caching and routing 
and show that an online gradient descent method can achieve a sublinear static regret. Considering competitive ratio, Chen \textit{et al.} \cite{chen2015online} proposes an online algorithm based on LASSO, while Lin \textit{et al.} \cite{lin2012online} and Shi \textit{et al.} \cite{shi2019value} modify receding horizon control algorithm. All these studies focus on comparison with static optimal policy. 

Dynamic regret is first introduced by Zinkevich \cite{zinkevich2003online}. Chen \textit{et al.} \cite{chen2017online} proposes an adaptive online saddle-point method and studies its dynamic regret. By allowing temporary constraint violation, Jin \textit{et al.} \cite{jin2020provisioning, jin2020resource} proposes different online learning models with a dynamic regret bound. However, these studies do not consider instantiating costs. 

Some recent studies explore using predictions to improve the performance of online algorithms. Considering precise request predictions, Chen \textit{et al.} \cite{chen2018smoothed} and Goel \textit{et al.} \cite{goel2019online} study an online caching problem with 2-norm instantiating costs and propose different algorithms with low competitive ratios. In addition, Comden \textit{et al.} \cite{comden2019online} and Li \textit{et al.} \cite{li2020online} propose online caching algorithms and analyze their dynamic regret. Furthermore, Chen \textit{et al.} \cite{chen2016using} and Li \textit{et al.} \cite{li2020leveraging} consider noisy predictions and analyze dynamic regret of their proposed algorithms. These studies, however, do not guarantee to produce integer solutions, and hence are not applicable to service caching when the services are indivisible.

\section{System Model}
\label{section:model}

We consider a system with multiple clients, an edge server, and a remote data center providing $N$ different services. The edge server is located near the clients and can cache a small subset of services. Any request from clients sent to the server can be processed immediately if the corresponding service is cached locally, 
otherwise it is forwarded to the remote center for processing.

Assume that time is slotted, and the total number of time slots is $T$. The edge server can dynamically adjust the set of services it caches. However, changing the set of cached services involves time-consuming operations such as downloading and setting up new services. Hence, we assume that the edge server can only adjust its cached services at the beginning of each time slot.

Let $x_{n,t}\in\{0,1\}$ denote the caching decision for service $n$ at time $t$. Let $X_{t} :=[x_{1,t} , x_{2,t} ,\dots,x_{N,t} ]$ be the caching decision at time $t$ and $X_{a:b} := [X_a, X_{a+1},\dots,X_{b}]$. Since the edge server often has limited storage, we assume that at most $M$ services can be cached at any time, that is,
\begin{align}
    \sum_{n=1}^{N}x_{n,t} \leq M, \quad \forall t.
    \label{eq:ineq_M}
\end{align}


Whenever the edge server caches a new service, it needs to download and install the said service. We model the cost of downloading and installing service $n$ by imposing an \emph{instantiating cost} of $\beta_n$. Thus, the total instantiating cost at time $t$ is 
\begin{align*}
    \sum_{n=1}^{N}\beta_n|x_{n,t}-x_{n,t-1}|_+,
\end{align*}
where $|x|_+:=\max\{x, 0\}$ for any $x\in \mathbb{R}$.

Next, we discuss the model for request arrivals and processing. Denote the total number of requests for service $n$ in time slot $t$ as $\lambda_{n,t}$. Let $\Lambda_t=[\lambda_{1,t},\lambda_{2,t},\dots,\lambda_{N,t}]$ and $\Lambda_{a:b} := [\Lambda_a,\Lambda_{a+1},\dots,\Lambda_{b}]$. We make the following mild assumption about $\Lambda_t$: If service $n$ and service $m$ are both among the top $M+1$ most popular services at time $t$, then $\lambda_{n,t}\neq\lambda_{m,t}$. This mild assumption is to ensure that the ordering of the top $M$ services is always unique.

The edge server can process all requests for its cached services locally. For services not cached at the edge, i.e., $x_{n,t}=0$, the edge server must forward all associated requests to the remote data center for processing, which inevitably leads to larger delays. The round-trip time between the edge server and the remote data center is determined by the conditions of the backbone network and the remote data center, and is little impacted by the edge server's caching decisions. Hence, we assume that there is a constant delay for requests that are processed by the remote data center, and say that the system suffers a constant \emph{forwarding cost} of $\alpha$ for each forwarded request. The total forwarding cost in time slot $t$ is then $$\alpha\sum_{n=1}^{N}\lambda_{n,t}(1-x_{n,t}).$$

Therefore, the total cost in time slot $t$ can be written as
\begin{align*}
   F_t(X_t,X_{t-1}) := \!\sum_{n=1}^{N}( \alpha\lambda_{n,t}(1-x_{n,t}) + \beta_n|x_{n,t}-x_{n,t-1}|_+ ).
\end{align*}

The goal of the edge server is to solve the problem of minimizing the total cost, which is shown below.
\begin{align}
    \min_{X_{1:T}} \quad & \sum_{t=1}^{T}F_t(X_t,X_{t-1}), \label{eq:obj}\\
    \textrm{s.t.} \quad& x_{n,t}\in\{0,1\}, \quad \forall n, \forall t, \label{eq:const1}\\
    & \sum_{n=1}^{N}x_{n,t}\leq M, \quad  \forall t. \label{eq:const2}
\end{align}


Note that solving this problem exactly is already challenging in the offline setting (i.e., all request arrivals are known in advance) due to the binary constraint in (\ref{eq:const1}). It is even more so (if not impossible) in the online setting, where the edge server needs to determine caching decision $X_t$ at the beginning of each time slot $t$ given limited knowledge about future request arrivals. We assume that the edge server employs an online algorithm and has exact predictions of request arrivals only in next $W$ time slots at any time $t$. Note that setting $W=0$ would correspond to the case where the edge server has no prediction ability; the case of using imprecise predictions is left for future work.  The concept of an online algorithm is formally defined as follows:

\begin{definition}
An online service caching algorithm is one that, after knowing $X_{1:t-1}$ and $\Lambda_{1:t+W-1}$, determines, possibly at random, $X_t$ at time $t$.
\end{definition}

The expected cost of an online algorithm $\xi$ is denoted by $C(\xi):=E[\sum_{t=1}^T F_t(X_t, X_{t-1})|\xi]$, where $E[\cdot]$ denotes the expectation function over all possible randomness.

To measure the performance of $\xi$, we compare the total cost of algorithm $\xi$ to that of an optimal dynamic offline policy, which is formally defined as follows.
\begin{definition} [Optimal Dynamic Offline Policy (OPT)]
An optimal dynamic offline policy is one that produces optimal solution $X_{1:T}^*$ for the problem in  \eqref{eq:obj}--\eqref{eq:const2}.
\end{definition}

Note that we allow any offline algorithm to cache different services in different time slots. This feature makes our work different from most existing studies on service caching that only consider optimal static offline policies, where the same set of services is cached in all time slots.

The difference between the expected cost of $\xi$ and the cost of optimal dynamic offline policy, denoted by $C(OPT)$, is called expected dynamic regret, i.e.,
\begin{equation}
    Reg(\xi) := C(\xi)-C(OPT). \label{eq:regret}
\end{equation}

Obviously, the expected dynamic regret of any online policy depends on the request arrivals $\Lambda_{1:T}$. We characterize $\Lambda_{1:T}$ by its \emph{path length}. Specifically, let $\theta_{n,t}$ be the indicator function that service $n$ is among the top $M$ services with the most requests in time slot $t$. Then, the path length of $\Lambda_{1:T}$ is defined as $\sum_{t=1}^{T} \sum_{n=1}^N |\theta_{n,t}-\theta_{n,t-1}|$. Let $\Theta_t:=[\theta_{1,t},\theta_{2,t},\dots,\theta_{N,t}]$. Loosely speaking, the path length measures the variation of the request distribution over time. We assume that the path length of $\Lambda_{1:T}$ is upper-bounded by $H_T$, i.e., $\sum_{t=1}^{T}\|\Theta_t-\Theta_{t-1}\|_1\leq H_T$, and the edge server knows the value of $H_T$. 

The goal of this work is to develop an online service caching algorithm whose expected dynamic regret is $o(T)$ whenever $H_T=o(T)$.


\section{Randomized Online Service Caching Algorithm}
\label{sec:alg}

In this section, we propose a randomized online service caching algorithm. Our algorithm mainly consists of two components. The first component determines the probability of caching a service $n$ at time $t$ with the goal of minimizing an auxiliary cost function. The second component is a randomized algorithm that determines which service to be cached at the edge based on the result of the first component while limiting the resulting instantiating cost. As we will show in the next section, combining these two components gives rise to an upper bound on the expected dynamic regret.

To express the probability distribution of $X_t$, we construct $K$ sample paths, each representing a probability mass of $\frac{1}{K}$. At the beginning of the whole process, the edge server chooses a number $k^*$ uniformly at random from $\{1,2,\dots, K\}$. Then, it uses the sample path $k^*$ at time $t$ as the caching decision in time $t$. 

For sample paths designed above, the portion of sample paths that cache a service is the same as the probability we cache this service. Let $p_{n,t}$ be the probability that the edge server caches service $n$ at time slot $t$, and let $P_t:=[p_{1,t}, p_{2,t},\dots, p_{N,t}]$ and $P_{a:b} := [P_a, P_{a+1}, \dots, P_b]$. Due to (\ref{eq:ineq_M}), $P_t$ is restricted to be in the following feasible set
\begin{equation}
    \mathbb{D}:=\Big\{[p_1,\dots, p_N]~|~0\leq p_{n}\leq 1, \forall n, \sum_{n=1}^Np_{n}\leq M \Big\}. \label{eq_setD}
\end{equation}

For decisions in sample paths, we use $s_{k,n,t}\in\{0,1\}$ to denote the indicator function that service $n$ is cached on sample path $k$ at time $t$, and let $S_{k,t}:=[s_{k,1,t}, s_{k,2,t}, \dots]$. Then, the edge server sets $X_t=S_{k^*,t}$ in each time slot $t$ as caching decisions. Thus, a randomized online service caching algorithm is effectively one that determines $S_{1,t}, S_{2,t},\dots, S_{K,t}$, in each time slot $t$.

As described above, our algorithm consists of two parts in each time slot $t$. In particular, we first determine caching probability $P_t$ based on previous probabilities and request arrivals $\Lambda_{t-1:t+W-1}$. Then, we use $P_t$ and sample paths at $t-1$, i.e., $[S_{1,t-1},S_{2,t-1},\dots,S_{K,t-1}]$, to determine sample paths at $t$.  
The overall algorithm is shown in Algorithm~\ref{alg:ROSC} and detailed steps are given in the next subsections. Here, to simplify notation, we let our algorithm start from $t=-W+1$ with $\Lambda_t$, $S_{k,t}$, $P_t$ set to zero for all $t\leq 0$. 

\begin{algorithm}   
\caption{Randomized Online Service Caching (ROSC)}
\begin{algorithmic}[1]
\renewcommand{\algorithmicrequire}{\textbf{Parameter:}}
\REQUIRE $K$
\STATE Choose $k^*$ uniformly at random from $\{1,2,\dots,K\}$
\STATE $\bar{P}_{-W+1:T}\leftarrow \textbf{0}$
\FOR{$t=-W+1$ to $T$}
\STATE Obtain parameter $\Lambda_{t+W-1}$
\STATE Apply HeapSort on $\Lambda_{t+W-1}$ to calculate $\Theta_{t+W-1}$
\STATE $P_{t+W}\leftarrow \Theta_{t+W-1}$
\IF{$W>0$}
\STATE $P_{t:t+W-1},\bar{P}_{t:t+W-1}\scriptstyle\leftarrow\textstyle$\mbox{Algo.\! \ref{alg:OU}$(\Lambda_{t:t+W-1},P_{t-1:t+W},$} $\bar{P}_{t:t+W-1},t)$
\ENDIF
\IF{$t\geq1$}
\STATE $[S_{1,t},\dots,S_{K,t}]\leftarrow$ Algo.~\ref{alg:RSC}$(P_t,S_{1,t-1},\dots,S_{K,t-1})$
\STATE $X_t\leftarrow S_{k^*,t}$
\ENDIF
\ENDFOR
\end{algorithmic}
\label{alg:ROSC}
\end{algorithm}


\subsection{Caching Probability Update}
Let us now discuss in detail our approach for determining $P_t$ in the first part of our algorithm. Define the following auxiliary cost function $\hat{F}_t$, which will be used as our surrogate objective function. 
\begin{align}
    &\hat{F}_t(P_t,P_{t-1}):= \sum_{j\,:\,0\leq p_{j,t}-p_{j,t-1}\leq\gamma}\frac{3\beta_j}{\gamma}(p_{j,t}-p_{j,t-1})^2 + \nonumber \\
    & \sum_{i\,:\,p_{i,t}-p_{i,t-1}>\gamma}3\beta_i(p_{i,t}-p_{i,t-1}) + \alpha\sum_{1\le n\le N}\lambda_{n,t}(1-p_{n,t}),
    \label{eq:auxi}
\end{align}
where $\gamma>0$ is a parameter whose value will be discussed in the next section. By comparing $\hat{F}_t$ with $F_t$, one can see that the only difference is in the instantiating cost component. Here, the quadratic term is to ensure that $\hat{F}_t$ is differentiable everywhere, and a factor of $3$ is added in order to bound the expected dynamic regret introduced by the randomized algorithm that will be discussed in the next section. 

At each time $t$, after obtaining the prediction $\Lambda_{t+W-1}$, the edge server first sets $P_{t+W}=\Theta_{t+W-1}$, i.e., $p_{n,t+W}=1$ if service $n$ is among the top $M$ most requested services in time slot $t+W-1$, and $p_{n,t+W}=0$, otherwise. If $W>0$, we will further update $P_{t:t+W-1}$ so as to reduce $\sum_{\tau=t}^{t+W-1} \hat{F}_\tau(P_\tau, P_{\tau-1})$ through projected gradient descent with step size $\eta$. 

Note that each $P_\tau$ only appears in $\hat{F}_\tau$ and $\hat{F}_{\tau+1}$. Thus, we obtain the gradient of $\sum_{\tau=t}^{t+W-1} \hat{F}_\tau(P_\tau, P_{\tau-1})$ with respect to $P_\tau$, denoted as  $\nabla_{P_{\tau}}(\hat{F}_{\tau}(P_{\tau},P_{\tau-1})+\hat{F}_{\tau+1}(P_{\tau+1},P_{\tau}))$, where
\begin{align}
    &\frac{\partial}{\partial p_{n,\tau}} \big(\hat{F}_\tau(P_{\tau},P_{\tau-1}) + \hat{F}_{\tau+1}(P_{\tau+1},P_{\tau})\big) \label{eq:part_gradient}\\
    &=\begin{cases}
    g_n(p_{n,\tau-1},p_{n,\tau})-\alpha\lambda_{n,\tau} - g_n(p_{n,\tau},p_{n,\tau+1}) &\textbf{if } \tau<T \\
    g_n(p_{n,\tau-1},p_{n,\tau}) -\alpha\lambda_{n,\tau} &\textbf{if } \tau=T 
    \end{cases}\nonumber
\end{align}
and the value of $g_n(a,b)$ is set to be $0$ if $b-a<0$, set to be $\frac{6\beta_n}{\gamma}(b-a)$ if $0\leq b-a\leq\gamma$, and set to be $3\beta_n$ if $b-a>\gamma$.

Then, we update $P_\tau$ from $\tau=t+W-1$ down to $\tau=t$. To ensure the gradient of $\sum_{\tau=t}^{t+W-1} \hat{F}_\tau(P_\tau, P_{\tau-1})$ with respect to $P_\tau$ is obtained based on $P_{\tau-1}$, $P_\tau$, and $P_{\tau+1}$ with the same update times, we use the updated $P_{\tau+1}$, the original $P_\tau$ and the $P_{\tau-1}$ in the previous iteration before its update, which is denoted as $\bar{P}_{\tau-1}$, to calculate the gradient. Thus, we update $P_\tau$ by 
$$P_\tau=\Pi_\mathbb{D}(P_\tau - \eta \nabla_{P_{\tau}} (\hat{F}_{\tau}(P_{\tau},\bar{P}_{\tau-1})+\hat{F}_{\tau+1}(P_{\tau+1},P_{\tau})),$$
where $\Pi_\mathbb{D}(\cdot)$ is the projection operator onto set $\mathbb{D}$ given in~\eqref{eq_setD}. This distinction is important for establishing an expected dynamic regret bound, as will be discussed in Section~\ref{sec:DR}. Algorithm~\ref{alg:OU} shows the detail of updating $P_{t:t+W-1}$.

\begin{algorithm}   
\caption{Projected Gradient Descent}
\begin{algorithmic}[1]
\renewcommand{\algorithmicrequire}{\textbf{Input:}}
\REQUIRE $\Lambda_{t:t+W-1},P_{t-1:t+W},\bar{P}_{t-1:t+W},t$
\renewcommand{\algorithmicrequire}{\textbf{Parameter:}}
\REQUIRE $\gamma,\eta$
\FOR{$\tau=t+W-1$ to $\max\{1,t\}$}
\STATE Calculate $\nabla_{P_{\tau}}(\hat{F}_{\tau}(P_{\tau},\bar{P}_{\tau-1})+\hat{F}_{\tau}(P_{\tau+1},P_{\tau}))$ by (\ref{eq:part_gradient})
\STATE $\bar{P}_{\tau}\leftarrow P_{\tau}$
\STATE $P_{\tau}\scriptstyle\leftarrow \textstyle\Pi_\mathbb{D}(P_\tau - \eta \nabla_{P_{\tau}} (\hat{F}_{\tau}(P_{\tau},\bar{P}_{\tau-1})+\hat{F}_{\tau+1}(P_{\tau+1},P_{\tau}))$
\ENDFOR
\renewcommand{\algorithmicrequire}{\textbf{Output:}}
\REQUIRE $P_{t:t+W-1},\bar{P}_{t:t+W-1}$
\end{algorithmic}
\label{alg:OU}
\end{algorithm}

\subsection{Sample Path Update}
Our algorithm for determining $[S_{k,t}]$ employs that in Fan \textit{et al.} \cite{fan2021online}, which studies online randomized algorithm for a different setting without establishing expected dynamic regret bound. The first step is to quantize every $p_{n,t}$ in $P_t$ into a multiple of $\frac{1}{K}$, denoted as $p^Q_{n,t}$. Let $P^Q_{t}:=[p^Q_{1,t},\dots,p^Q_{N,t}]$. Note that each service $n$ in $[S_{k,t}]$ needs to be cached in exactly $Kp^Q_{n,t}$ sample paths. Set sample path $S_{k,t}=S_{k,t-1}$ for all $k$ at time $t$. Then, for each $n$, randomly choose $K(p^Q_{n,t}-p^Q_{n,t-1})$ sample paths without service $n$ to cache service $n$ if $p^Q_{n,t}>p^Q_{n,t-1}$, and delete service $n$ from $K(p^Q_{n,t-1}-p^Q_{n,t})$ randomly chosen sample paths with service $n$ if the $p^Q_{n,t}<p^Q_{n,t-1}$. Finally, for each sample path $k$ that caches more than $M$ services, find another sample path $k'$ with less than $M$ cached services, and randomly choose a service $n$ that $k$ caches and $k'$ does not. Delete service $n$ from $k$ and cache it in the $k'$. Detailed steps are shown in Algorithm~\ref{alg:RSC}. This algorithm is designed so that the number of changes, which corresponds to the instantiating cost at time $t$, can be bounded.

\begin{algorithm} 
\caption{Randomized Caching}
\begin{algorithmic}[1]
\renewcommand{\algorithmicrequire}{\textbf{Input:}}
\REQUIRE $P_t,S_{1,t-1},S_{2,t-1},\dots,S_{K,t-1}$
\STATE $P_t^Q\leftarrow$ quantize every $p_{n,t}$ in $P_t$ into a multiple of $\frac{1}{K}$
\STATE $P_{t-1}^Q\leftarrow \frac{1}{K}\sum_{k=1}^{K}S_{k,t-1}$
\STATE $\Delta_t:=[\delta_{1,t},\dots,\delta_{N,t}]\leftarrow P_t^Q-P_{t-1}^Q$
\STATE $S_{1,t},S_{2,t},\dots,S_{K,t}\leftarrow S_{1,t-1},S_{2,t-1},\dots,S_{K,t-1}$
\FOR{$n=1,2,\dots,N$}
\IF{$\delta_{n,t}>0$}
\STATE Find the set $\{s_{k,n,t}|s_{k,n,t}=0\}$, randomly pick $K\delta_{n,t}$ elements in it and set to $1$ 
\ELSIF{$\delta_{n,t}<0$}
\STATE Find the set $\{s_{k,n,t}|s_{k,n,t}=1\}$, randomly pick $|K\delta_{n,t}|$ elements in it and set to $0$ 
\ENDIF
\ENDFOR
\WHILE{$\exists \sum_{n=1}^{N}s_{k,n,t} > M$}
\STATE Find $k'$ that $\sum_{n=1}^{N}s_{k',n,t}< M$
\STATE Randomly choose a service $n'$ from the set $\{n|s_{k',n,t}=0,s_{k,n,t}=1\}$
\STATE $s_{k',n',t}\leftarrow 1$, $s_{k,n',t}\leftarrow 0$
\ENDWHILE
\renewcommand{\algorithmicrequire}{\textbf{Output:}}
\REQUIRE $S_{1,t},S_{2,t},\dots,S_{K,t}$
\end{algorithmic}
\label{alg:RSC}
\end{algorithm}

\section{Expected Dynamic Regret}
\label{sec:DR}
In this section, we analyze the regret of ROSC. The main result is the following.
\begin{theorem}
    Let $\gamma=\sqrt{\frac{H_T}{T}}$ and $\eta=\frac{\gamma}{12\beta^*}$ with $\beta^*:=\max_{n}\beta_n$. If the number of requests in each time slot is upper-bounded by $U$, that is, $\sum_{n=1}^{N}\lambda_{n,t}\leq U, \; \forall t$, then 
    \begin{align}
        Reg(ROSC) &\leq \Big(\frac{6\sqrt{2M}\beta^*(\alpha+3\beta^*)}{\alpha W}+ 3\beta^* N \Big) \sqrt{H_T T} \nonumber\\ 
        &\quad + \frac{(\alpha U + 6\beta^*N) T}{K} + 2\beta^* H_T.\label{eq_Reg_ROSC}
    \end{align}
    In particular, $Reg(ROSC)=o(T)$ if $H_T=o(T)$ and $K=\sqrt{T}$.
    \label{theorem:regret} 
\end{theorem}

We will prove this result in two steps. First, let $P_{1:T}^{'}$ be the final value of $P_{1:T}$ in Algorithm~\ref{alg:ROSC} and let $P_{1:T}^*$ be the optimal vector for minimizing the  auxiliary cost function $\sum_{t=1}^{T}\hat{F}_t(P_t, P_{t-1})$ under the constraint (\ref{eq:const2}). We will derive an upper bound on $\sum_{t=1}^{T}\hat{F}_t(P_t^{'}, P_{t-1}^{'})-\sum_{t=1}^{T}\hat{F}_t(P_t^*, P_{t-1}^*)$. Second, we will show that $Reg(ROSC)$, which is defined with respect to $F_t(\cdot)$ instead of $\hat{F}_t(\cdot)$, can actually be bounded by a function of $\sum_{t=1}^{T}\hat{F}_t(P_t^{'}, P_{t-1}^{'})-\sum_{t=1}^{T}\hat{F}_t(P_t^*, P_{t-1}^*)$.

\subsection{Bounding $\sum_{t=1}^{T}\hat{F}_t(P_t^{'}, P_{t-1}^{'})-\sum_{t=1}^{T}\hat{F}_t(P_t^*, P_{t-1}^*)$}

We first compare ROSC with an offline policy and then bound $\sum_{t=1}^{T}\hat{F}_t(P_t^{'}, P_{t-1}^{'})-\sum_{t=1}^{T}\hat{F}_t(P_t^*, P_{t-1}^*)$. 
%
Consider an offline policy that knows $\Lambda_{1:T}$ and employs the projected gradient descent algorithm to minimize
\begin{align*}
    J(Q) := \sum_{t=1}^{T}\hat{F}_t(Q_t,Q_{t-1})
\end{align*}
subject to the constraint  $Q=[Q_1,\dots,Q_T] \in \mathbb{H}$, where $Q_t:=[q_{1,t},q_{2,t},\dots,q_{N,t}]$ and $\mathbb{H} := \{ Q ~|~ 0\leq q_{n,t}\leq 1, \forall n,t, \sum_{n=1}^N q_{n,t}\leq M,\forall t\}$. Following a projected gradient descent algorithm, the offline policy first initializes $Q_t^0=\Lambda_{t-1}$ and then updates its caching decisions $Q$ in each iteration $w = 1,\ldots,W$ as follows 
\begin{align}
Q^w \leftarrow \Pi_\mathbb{H}\Big( Q^{w-1}  -  \eta\nabla J(Q^{w-1})\Big).\label{eq_update_Q}
\end{align}
Note that the following has been shown in Li \textit{et al.} \cite{li2020online}.
\begin{lemma}
For update \eqref{eq_update_Q}, we have
    $Q_t^W=P_t^{'},\forall t$.
    \label{lemma:equality}
\end{lemma}
Using this result, we can prove the following. 
\begin{lemma}
    Consider ROSC with step size  $\eta=\frac{\gamma}{12\beta^*}$. Then
    \begin{align*}
         &\sum_{t=1}^{T}\hat{F}_t(P_t^{'},P_{t-1}^{'}) -\sum_{t=1}^{T} \hat{F}_t(P_t^*,P_{t-1}^*) \\
        &\leq \frac{6\beta^*}{\gamma W}\sum_{t=1}^{T}\|\Theta_{t-1}-P_t^*\|_2^2.
    \end{align*}
    \label{lemma:regret_OFGD}
\end{lemma}
\begin{IEEEproof}
First, it can be seen that $J(\cdot)$ is $\frac{12\beta^*}{\gamma}$ smooth. 
Then the result follows by simply applying  \cite[Theorem~10.21]{beck2017first} to the offline policy~\eqref{eq_update_Q} and then using Lemma~\ref{lemma:equality}. 
\end{IEEEproof}

Next, we bound the term  $\sum_{t=1}^{T}\|\Theta_{t-1}-P_t^*\|_2^2$. In fact,
\begin{lemma} 
We have
    \begin{align}
        \sum_{t=1}^{T}\|\Theta_{t-1}-P_t^*\|_2^2\leq\frac{\sqrt{2M}(\alpha+3\beta^*)}{\alpha}H_T.
    \end{align}
    \label{lemma:MFD}
\end{lemma}
\begin{IEEEproof}
    First, note that if $0\leq p_{j,t}-p_{j,t-1}\leq\gamma$, then $\frac{3\beta_j}{\gamma}(p_{j,t}-p_{j,t-1})^2 \le 3\beta_j(p_{j,t}-p_{j,t-1}).$ 
    Using this and the definitions of $\hat{F}_t$, we have $\hat{F}_t(\Theta_t, \Theta_{t-1})\leq \alpha\sum_{n=1}^{N}\lambda_{n,t}(1-\theta_{n,t}) + 3\sum_{n=1}^{N}\beta_n |\theta_{n,t}-\theta_{n,t-1}|_{+}$.

    Since $P^*_{1:T}$ 
    minimizes $\sum_{t=1}^{T}\hat{F}_t(P_t,P_{t-1})$, we have $\sum_{t=1}^{T}\hat{F}_t(P^*_t,P^*_{t-1}) \leq \sum_{t=1}^{T}\hat{F}_t(\Theta_t, \Theta_{t-1})\leq\sum_{t=1}^{T}\sum_{n=1}^N \Big(\alpha\lambda_{n,t}(1 - \theta_{n,t})+3\beta_n |\theta_{n,t}-\theta_{n,t-1}|_{+})\Big)$.
    Plugging in the definition of $\hat{F}_t(P^*_t,P^*_{t-1})$ and then rearranging this relation yields $\alpha\sum_{t=1}^{T}\sum_{n-1}^{N}\lambda_{n,t}(\theta_{n,t}-p^*_{n,t}) \leq 3\sum_{t=1}^{T}\sum_{n=1}^{T}\beta_n |\theta_{n,t}-\theta_{n,t-1}|_{+} \leq 3\beta^* H_T$.

    Without loss of generality, we can assume that  $\lambda_{1,t}\geq\lambda_{2,t}\geq\dots\geq\lambda_{N,t}$ for a given $t$. Then, $\lambda_{n,t}\geq\lambda_{n+1,t}+1$ for $1\leq n\leq M$. Combining this with the fact that $\theta_{1,t}=\theta_{2,t}=\dots=\theta_{M,t}=1$ and $\theta_{M+1,t}=\theta_{M+2,t}=\dots=\theta_{N,t}=0$, we have $\sum_{n=1}^{N}\lambda_{n,t}\theta_{n,t}-\sum_{n=1}^{N}\lambda_{n,t}p_{n,t}^{*}\geq \sum_{n=1}^{N}|\theta_{n,t}-p_{n,t}^*|.$
    Therefore,
    \begin{align*}
        \sum_{t=1}^{T}\sum_{n=1}^{N}|\theta_{n,t}-p_{n,t}^*|\leq\sum_{t=1}^{T}\sum_{n=1}^{N}\lambda_{n,t}(\theta_{n,t}-p_t^{*}) \leq \frac{3\beta^* H_T}{\alpha}.
    \end{align*}
    
    Next, using the triangle inequality, we have
    \begin{align*}
        & \sum_{t=1}^{T}||\Theta_{t-1}-P_t^{*}||_2  \leq \sum_{t=1}^{T}||\Theta_{t-1}-\Theta_t||_2 + \sum_{t=1}^{T}||\Theta_t-P_t^{*}||_2 \\
        & \leq  \sum_{t=1}^{T}||\Theta_{t-1}-\Theta_t||_1 + \sum_{t=1}^{T}||\Theta_t-P_t^{*}||_1 \\
        & \leq H_T + \frac{3\beta^* H_T}{\alpha} = \frac{\alpha+3\beta^*}{\alpha}H_T.
    \end{align*}
    Since the caching limit is $M$, it follows that $\|\Theta_{t-1}-P_t^{*}\|_2\leq \sqrt{2M}$. As a result, 
    \begin{align*}
        \sum_{t=1}^{T}||\Theta_{t-1}-P_t^{*}||_2^2 &\leq \sqrt{2M}\sum_{t=1}^{T}||\Theta_{t-1}-P_t^{*}||_2\\
        & \leq \frac{\sqrt{2M}(\alpha+3\beta^*)}{\alpha}H_T.
    \end{align*}
This completes the proof of the lemma.
\end{IEEEproof}

Now, by combining  Lemma~\ref{lemma:MFD} and Lemma~\ref{lemma:regret_OFGD}, we obtain
\begin{align}
    &\sum_{t=1}^{T}\hat{F}_t(P_t^{'},P_{t-1}^{'})-\sum_{t=1}^{T}\hat{F}_t(P_t^*,P_{t-1}^*) \nonumber\\
    & \leq \frac{6\sqrt{2M}\beta^*(\alpha+3\beta^*)}{\alpha\gamma W}H_T.
    \label{eq:auxi_regret}
\end{align}

\subsection{Bounding $Reg(ROSC)$}

We now analyze the cost introduced by the auxiliary objective function and the randomized algorithm, and then bound $Reg(ROSC)$. 

Considering the structure of the auxiliary cost function and the analysis of the randomized algorithm in \cite{fan2021online}, we can show the following.
\begin{lemma}
    By choosing $0<\gamma<1$,
    \begin{align}
        Reg&(ROSC)  \leq \sum_{t=1}^{T}\hat{F}_t(P_t^{'},P_{t-1}^{'})-\sum_{t=1}^{T}\hat{F}_t(P_t^*,P_{t-1}^*) \nonumber\\
        & + 3\gamma\beta^* N T + \frac{(\alpha U + 6\beta^*N) T}{K} + 2\beta^* H_T.
    \end{align}
    \label{lemma:auxi}
\end{lemma}
\begin{IEEEproof}
    It has been shown in \cite{fan2021online} that, under ROSC, $E[x_{n,t}]=p_{n,t}^Q$ and $E\big[\sum_{t=1}^{T}\sum_{n=1}^{N}|x_{n,t}-x_{n,t-1}|_+\big]\leq 3\sum_{t=1}^{T}\sum_{n=1}^{N}|p_{n,t}^Q-p_{n,t-1}^Q|_+$, where $p_{n,t}^Q$ is the quantized version of $p_{n,t}^{'}$. Hence, we have
    \begin{align*}
        E[\sum_{t=1}^{T}F_t(X_t,X_{t-1})] &\leq \sum_{t=1}^{T}\sum_{n=1}^{N}\alpha\lambda_{n,t}(1-p_{n,t}^Q) \\
        & +  3\sum_{t=1}^{T}\sum_{n=1}^{N}\beta_{n}|p_{n,t}^Q-p_{n,t-1}^Q|_+.
    \end{align*}
    Since the difference between $p_{n,t}^Q$ and $p_{n,t}^{'}$ is at most $\frac{1}{K}$ according to the design of Algorithm~\ref{alg:RSC}, we have
        \begin{align*}
        &E[\sum_{t=1}^{T}F_t(X_t,X_{t-1})] \leq \sum_{t=1}^{T}\sum_{n=1}^{N}\alpha\lambda_{n,t}(1-p_{n,t}^{'}) + \frac{\alpha UT}{K}\\
        & \; + 3\sum_{t=1}^{T}\sum_{n=1}^{N}\beta_{n}|p_{n,t}^{'}-p_{n,t-1}^{'}|_+ + \frac{6\beta^*NT}{K} \\
        & \leq \sum_{t=1}^{T}\hat{F}_t(P_t^{'},P_{t-1}^{'}) + 3\beta^* \gamma N T + \frac{(\alpha U + 6\beta^*N) T}{K}.
    \end{align*}

    Then, by comparing $\hat{F}_t(\cdot)$ and $F_t(\cdot)$, we have
    \begin{align*}
        &C(OPT)  = \sum_{t=1}^{T}F_t(X_t^*, X_{t-1}^*) \\
        & \geq \sum_{t=1}^{T}\hat{F}_t(X_t^*, X_{t-1}^*) - 2\sum_{t=1}^{T}\sum_{n=1}^{N}\beta_{n}|x_{n,t}^*-x_{n,t-1}^*|.
    \end{align*}

    Thus,
    \begin{align*}
        & Reg(ROSC) = E[\sum_{t=1}^{T}F_t(X_t,X_{t-1})] - C(OPT)\\
        & \leq \sum_{t=1}^{T}\hat{F}_t(P_t^{'},P_{t-1}^{'}) + 3\gamma\beta^* N T + \frac{(\alpha U + 6\beta^*N) T}{K} \\
        & \quad - C(OPT) \\
        & \leq \sum_{t=1}^{T}\hat{F}_t(P_t^{'},P_{t-1}^{'}) - \sum_{t=1}^{T}\hat{F}_t(X_t^*,X_{t-1}^*) + 3\gamma\beta^* N T \\
        & \quad + 2\sum_{t=1}^{T}\sum_{n=1}^{N}\beta_{n}|x_{n,t}^*-x_{n,t-1}^*|  + \frac{(\alpha U + 6\beta^*N) T}{K}\\
        & \leq \sum_{t=1}^{T}\hat{F}_t(P_t^{'},P_{t-1}^{'}) - \sum_{t=1}^{T}\hat{F}_t(P_t^*,P_{t-1}^*) + 3\gamma\beta^* N T \\
        & \quad + 2\sum_{t=1}^{T}\sum_{n=1}^{N}\beta_{n}|x_{n,t}^*-x_{n,t-1}^*|  + \frac{(\alpha U + 6\beta^*N) T}{K}.
    \end{align*}
    
    Note from the definitions of $\Theta_t$ and $X^*_{1:T}$ that 
    \begin{align*}
        \sum_{t=1}^{T}\sum_{n=t}^{N}(x_{n,t}^*-x_{n,t-1}^*) \leq \sum_{t=1}^{T}\sum_{n=t}^{N}(\theta_{n,t}-\theta_{n,t-1}).
    \end{align*}
    Therefore,
    \begin{align*}
        &Reg(ROSC) \leq \sum_{t=1}^{T}\hat{F}_t(P_t^{'},P_{t-1}^{'}) - \sum_{t=1}^{T}\hat{F}_t(P_t^*,P_{t-1}^*) \\
        & + 3\gamma\beta^* N T + \frac{(\alpha U + 6\beta^*) NT}{K} + 2\sum_{t=1}^{T}\beta_{n}\|\Theta_{n,t}-\Theta_{n,t-1}\|_1 \\
        & \leq \sum_{t=1}^{T}\hat{F}_t(P_t^{'},P_{t-1}^{'}) - \sum_{t=1}^{T}\hat{F}_t(P_t^*,P_{t-1}^*) + 3\gamma\beta^* N T \\
        & + \frac{(\alpha U + 6\beta^*N) T}{K} + 2\beta^* H_T .
    \end{align*}
    This completes the proof of the lemma. 
\end{IEEEproof}

We are now ready to prove Theorem~\ref{theorem:regret}.

\begin{IEEEproof}[Proof of Theorem \ref{theorem:regret}] 
By combining Lemma~\ref{lemma:auxi} and (\ref{eq:auxi_regret}), the expected dynamic regret is bounded by
\begin{align*}
    Reg(ROSC) \leq & \frac{6\sqrt{2M}\beta^*(\alpha+3\beta^*)}{\alpha\gamma W}H_T + 3\gamma\beta^* N T\\
    & + \frac{(\alpha U + 6\beta^*N) T}{K} + 2\beta^* H_T.
\end{align*}
By taking $\gamma=\sqrt{\frac{H_T}{T}}$, we obtain \eqref{eq_Reg_ROSC} as desired. 
\end{IEEEproof}


\section{An Efficient Implementation for ROSC}
\label{sec:Com}

In this section, we propose a projection algorithm to efficiently implement ROSC and then analyze the complexity of ROSC. The main result is shown below.
\begin{theorem}
    Using Algorithm~\ref{alg:proj} below for projection, the complexity of ROSC is $O(\max\{WN\log(N), KMN\})$ per time slot. 
    \label{theorem:comp}
\end{theorem}

An important bottleneck of the complexity when implementing ROSC is the projection step in step 4 of Algorithm~\ref{alg:OU}. In previous works, Wang \cite{wang2015projection} proposes an $O(N^2)$ algorithm for computing exact projections, and Beck \textit{et al.} \cite[p.~150]{beck2017first} demonstrates an algorithm based on a bisection method for computing an approximate projection onto a bounded simplex. Based on these ideas, we develop an efficient $O(N\log(N))$ projection algorithm for computing \emph{exact} projection onto the set $\mathbb{D}$ in Algorithm~\ref{alg:OU}. That is, given $Z\in \mathbb{R}^N$, find $ Y = \Pi_\mathbb{D}(Z).$ The idea of our projection algorithm is based on the following lemma.
\begin{lemma}
    If $Z$ is sorted in a descending order and $Y = \Pi_\mathbb{D}(Z)$, then $Y$ is also sorted in the same fashion, and there exists an index $i^* \in [0,N]$ such that $Y_{1:i^*}=\mathbf{1}$ and $Y_{(i^*+1):N} < \mathbf{1}$ is the projection of $Z_{(i^*+1):N}$ onto the simplex $\mathcal{S}_{i^*} = \{V\in [0,\infty)^{N-i^*}~|~\sum_{j=1}^{N-i^*} v_{j}=M-i^*\}$.
\end{lemma}
\begin{IEEEproof}
First, it is clear that $y_i=0$ if $z_i\leq 0$. Thus, $Y=\Pi_\mathbb{D}([Z]^{+})$ where $[Z]^{+}=\max \{Z,\mathbf{0} \}$. Moreover, if the projection of $Z$ onto $[0,1]^N$, denoted by $Y^{'}=\Pi_{[0,1]^N}(Z)$, is such that $\langle \mathbf{1}, Y^{'}\rangle \leq M$, then $Y=Y^{'}$. Thus, w.l.o.g., we will consider
\begin{align}
    Z \geq \mathbf{0}, \quad \langle \mathbf{1}, \Pi_{[0,1]^N}(Z) \rangle \geq M.
    \label{eq:proj_relax}
\end{align}
A consequence of (\ref{eq:proj_relax}) is that $\langle \mathbf{1}, Z \rangle \geq M$ and $\langle \mathbf{1}, Y \rangle = M$. Thus, we instead consider the following problem:
\begin{align}
    Y = \argmin_{Y\in [0,1]^N} \big\{\frac{1}{2}\|Z-Y\|_2^2~|~\langle \mathbf{1}, Y \rangle = M\big\}
    \label{eq:proj_robject}
\end{align}

Let us introduce a Lagrangian of (\ref{eq:proj_robject})
\begin{align*}
    L(Y,\mu,\nu,\rho) &= \frac{1}{2}\|Z-Y\|_2^2+\langle \nu,Y-\mathbf{1}\rangle-\langle\mu,Y\rangle \\
    &\quad +\rho(\langle\mathbf{1},Y\rangle-M),
\end{align*}
where $\mu,\nu,\rho$ are the corresponding Lagrange multipliers. Since the problem is convex, the KKT conditions are necessary and sufficient for optimality, i.e.,
\begin{align}
    y_i-z_i-\mu_i+\nu_i+\rho = 0, \forall i \label{eq:KKT1}& \\
    \mu_i y_i=0,\quad \nu_i(y_i-1)=0, \forall i \label{eq:KKT2}& \\
    0\leq y_i\leq 1, \quad \textstyle \sum_{i=1}^{N}y_i = M \label{eq:KKT3}& \\
    \mu\geq \mathbf{0},\quad \nu\geq \mathbf{0}, \quad \rho\in \mathbb{R}. \label{eq:KKT4}&
\end{align}

Clearly, if $0\leq y_i \leq 1$, then it must hold that $y_i=z_i-\rho$. As a result, the optimal solution can be partitioned as:
\begin{align*}
    \mathcal{I}_1=\{i|y_i=\mathbf{1}\},\mathcal{I}_2=\{i|y_i=z_i-\rho\},\mathcal{I}_3=\{i|y_i=\mathbf{0}\}.
\end{align*}

Since $M=\sum_{i=1}^{N}y_i=|\mathcal{I}_1|+\sum_{\mathcal{I}_2}(x_i-\rho)$, we have
\begin{align*}
    \rho |\mathcal{I}_2|=\sum_{i\in \mathcal{I}_2}z_i-(M-|\mathcal{I}_1|).
\end{align*}

Next, observe that
\begin{itemize}
    \item On $\mathcal{I}_1$: $\mu_i=0$ and $z_i=\mu_i+\rho+1\geq \rho + 1$.
    \item On $\mathcal{I}_2$: $\mu_i=\nu_i=0$ and $\rho<z_i<\rho+1$.
    \item On $\mathcal{I}_3$: $\nu_i=0$ and $z_i=\rho-y_i\leq \rho$.
\end{itemize}

The above facts imply that if $Z$ is sorted decreasing, then $Y$ is also sorted decreasing and can be expressed as
\begin{align*}
    Y=[\mathbf{1}_{1:i^*},\bar{Y}]
\end{align*}
where $i^*=|\mathcal{I}_1|$ and
\begin{align}
    \bar{Y}=[z_{(i^*+1):(i^*+|\mathcal{I}_2|)}-\rho, \mathbf{0}_{(i^*+|\mathcal{I}_2|+1):N}] < \mathbf{1}.
\end{align}

Assume $Z$ is sorted decreasing and $\hat{Z}:=[z_{i^*+1},\dots,z_N]$. Then, 
the projection of $\hat{Z}$ onto the simplex $\mathcal{S}_{i^*}$ is given by
\begin{align}
    \Tilde{Y}=\arg\min_{\Tilde{Y}\in \mathcal{S}}\big\{\frac{1}{2}\|\hat{Z}-\Tilde{Y}\|_2^2~|~\langle\mathbf{1},\Tilde{Y}\rangle = M-i^{*}\big\}.
    \label{eq:proj_s}
\end{align}
It is easy to verify that by using $(Y,\mu,\nu,\rho)$ satisfying (\ref{eq:KKT1})-(\ref{eq:KKT4}), $(\bar{Y},\{\nu_i\}_{i\geq i^*},\rho)$ satisfy the KKT conditions of problem (\ref{eq:proj_s}), and hence $\bar{Y}$ is the projection of $\hat{Z}$ onto simplex $\mathcal{S}_{i^*}$.

\end{IEEEproof}

By using this lemma, we can further show that $i^*$ is indeed the smallest index $i \in [0,N]$ such that the projection of $Z_{(i+1):N}$ onto the simplex $\mathcal{S}_{i}$ is strictly less than $1$; the proof is straightforward and thus skipped for brevity. 
As a result, when $Z$ is sorted in a descending order, we can use a binary search to find the index  $i^*$. Note that in each step of the search, we need to find the projection onto a simplex, which can be computed efficiently, e.g., using the algorithm in \cite{duchi2008efficient}. We recall this algorithm below.




\begin{algorithm}
\caption{$\Pi_{\sf simplex}(A,c)$: Projection onto a Simplex}
\begin{algorithmic}[1]
\renewcommand{\algorithmicrequire}{\textbf{Input:}}
\REQUIRE $A \in \mathbb{R}^m, c>0$ s.t. $a_1\geq a_2\geq\dots\geq a_m$
\STATE $I\leftarrow\max_{i\geq 1}\{i~|~(\sum_{j=1}^{m}a_j-c)/i<a_i$
\STATE $\tau\leftarrow (\sum_{j=1}^{m}a_j-c)/I$
\FOR{$j=1$ to $m$}
\STATE $a_j^*\leftarrow \max\{a_j-\tau, 0\}$
\ENDFOR
\renewcommand{\algorithmicrequire}{\textbf{Output:}}
\REQUIRE $A^*$
\end{algorithmic}
\label{alg:proj2}
\end{algorithm}
The runtime of Algorithm~\ref{alg:proj2} is linear in the input size. Therefore, by using a binary search and applying Algorithm~\ref{alg:proj2} repeatedly, we can find index $i^*$ in nearly linear time; 
the details are given in Algorithm~\ref{alg:proj} below. 

\begin{algorithm}
\caption{$\Pi_\mathbb{D}(Z)$: Projection onto a Bounded Simplex}
\begin{algorithmic}[1]
\renewcommand{\algorithmicrequire}{\textbf{Input:}}
\REQUIRE $Z \in \mathbb{R}^N, M>0$
\STATE $Z\leftarrow \max\{Z, \mathbf{0} \}$
\STATE $V\leftarrow \min\{Z, \mathbf{1} \}$
\IF{$\langle V,\mathbf{1}\rangle \leq M$}
\STATE $Y\leftarrow V$
\ELSE
\STATE $[Z,\mathtt{Id}]\leftarrow \mathsf{sort}(Z,\mathtt{'descend'})$
\STATE $V\leftarrow \mathbf{0},l\leftarrow0,r\leftarrow M$
\FOR{$n=0$ to $\lceil \log_2(M) \rceil$}
\STATE $i^*\leftarrow \lfloor (r+l)/2 \rfloor$
\STATE $Y^{'}\leftarrow \Pi_{\sf simplex} (Z_{(i^*+1):N},M-i^*)$
\IF{$i^*==l$}
\IF{any $y_i^{'} \geq 1$}
\STATE $V\leftarrow[\mathbf{1}_{1:r}, \Pi_{\sf simplex}(Z_{(r+1):N},M-r)]$
\ELSE
\STATE $V=[\mathbf{1}_{1:l},Y^{'}]$
\ENDIF
\STATE \textbf{break}
\ENDIF
\IF{any $y_i^{'} \geq 1$}
\STATE $l\leftarrow i^*$
\ELSE
\STATE $r\leftarrow i^*$
\ENDIF
\ENDFOR
\STATE $Y(\mathtt{Id})\leftarrow V$
\ENDIF
\renewcommand{\algorithmicrequire}{\textbf{Output:}}
\REQUIRE $Y$
\end{algorithmic}
\label{alg:proj}
\end{algorithm}

We now show that Algorithm~\ref{alg:proj} has low complexity.
\begin{lemma}
    By using HeapSort as the sorting method, the time complexity of Algorithm~\ref{alg:proj} is $O(N\log N )$.
    \label{lemma:pro_comp}
\end{lemma}
\begin{IEEEproof}
We analyze the time complexity of Algorithm~\ref{alg:proj} line by line. 
First, the complexity of lines 1--4 is $O(N)$. Then, the sorting operation in line 6 can be finished in $O(N \log N )$ using HeapSort. Finally, the loop in binary search runs at most $\log M$ times, each of which calls Algorithm~\ref{alg:proj2} once and thus takes only $O(N)$. 
Therefore, the overall time complexity of Algorithm~\ref{alg:proj} is $O(N \log N )$.
\end{IEEEproof}

We are ready to prove Theorem~\ref{theorem:comp}.

\begin{IEEEproof}[Proof of Theorem \ref{theorem:comp}]
In each time slot, ROSC's procedures include a single run of initialization, Algorithm~\ref{alg:OU}, Algorithm~\ref{alg:RSC} and assignment of $X_t$.

We first analyze the time complexity of Algorithm~\ref{alg:OU}. According to (\ref{eq:part_gradient}) and Lemma~\ref{lemma:pro_comp}, line 2, 3 and 4 in Algorithm~\ref{alg:OU} run in $O(N)$, $O(N)$ and $O(N\log N )$, respectively. Since the for-loop in Algorithm~\ref{alg:OU} runs at most $W$ times, the complexity of Algorithm~\ref{alg:OU} is $O(WN \log(N))$.

Next, Fan \textit{et al.} \cite{fan2021online} shows that the complexity of Algorithm~\ref{alg:RSC} is $O(KMN)$. For initialization and assignment in ROSC, it is easy to verify that the complexity is $O(N)$.

Therefore, the total complexity of ROSC per time slot is $O(\max\{WN\log(N), KMN\})$.
\end{IEEEproof}


\section{Evaluation}
\label{sec:simulation}

In this section, we evaluate the performance of ROSC through various simulations and compare it to that of other state-of-the-art policies. We also evaluate the case when the prediction of future arrivals can be inaccurate.


\subsection{Setup}

\textbf{Data.} We conduct experiments on two different data sets. The first data set is based on a random replacement model presented by Elayoubi \textit{et al.} \cite{elayoubi2015performance}. The requests in this data set follow a Zipf distribution, while the ranking of services changes frequently according to real-world measured statistics. We call this the \emph{Replacement data set}. The second data set follows the model introduced by Traverso \textit{et al.} \cite{traverso2013temporal}. Services are divided into 5 groups in which services share the same lifetime in the same group. The beginnings of the services follow a Poisson process determined by their group. We call this the \emph{Poisson data set}. Table~\ref{tab1} summarizes important parameters of data sets.

\begin{table}[htbp]
\caption{Request Model Parameters}
\begin{center}
\begin{tabular}{|c|c|c|c|c|c|}
\hline
Model & \textbf{\textit{$N$}}& \textbf{\textit{$T$}}& \textbf{\textit{$U$}} & \textit{Ranking Lifetime$^{\mathrm{*}}$} \\
\hline
Replacement& $10^3$ & $10^4$ & $200$ & Follow Table 2 in \cite{elayoubi2015performance} \\
\hline
Poisson& $10^3$ & $10^4$ & \multicolumn{2}{c|}{Follow Trace 1 in \cite{traverso2013temporal}} \\
\hline
\multicolumn{5}{l}{$^{\mathrm{*}}$ Represent how often the popularity of each service changes}
\end{tabular}
\label{tab1}
\end{center}
\vspace{-3mm}
\end{table}

\textbf{Default parameters.} Throughout the evaluation, we set $K=100$ for ROSC and assume $\beta_1=\beta_2=\dots=\beta_N=\beta^*$. Since the forwarding cost and instantiating cost per service vary for different edge servers, we fix $\alpha=0.05$ and then evaluate the total cost using different $\frac{\beta^*}{\alpha}$. For the auxiliary function, we set $\gamma=0.05$ as $T$ and $H_T$ are not available to the online algorithm. We also set the step size to be $\eta=\frac{\gamma}{12\beta^*}$ as suggested in Theorem~\ref{theorem:regret}.

\textbf{Comparison schemes.} We compare ROSC with four other algorithms:
\begin{itemize}
    \item Receding Horizon Control (RHC): RHC is introduced in \cite{comden2019online, camacho2013model, garcia1989model}. In each time slot $t$, it chooses to cache $X_t$ by solving the optimization problem $\argmin_{X_{t:t+W-1}}\sum_{\tau=t}^{t+W-1}F_\tau(X_\tau,X_{\tau-1})$.
    \item Committed Horizon Control (CHC): CHC is generalized RHC and has been proposed in \cite{comden2019online,chen2016using}. It's caching decision in time slot $t$ is the average of RHC solutions $X_t$ in the previous $W$ time slots. 
    \item Static Optimal Offline Algorithm (SOPT): This is an offline policy that has knowledge of all future requests and caches the same services in all time slots that minimize $\sum_{t=1}^{T}F_t(X_t,X_{t-1})$. Specifically, it caches the same $M$ services with the largest total requests with $\sum_{t=1}^{T}\lambda_{n,t}\geq \frac{\beta^*}{\alpha}$ in all time slots.
    \item ROSC, W=300: Lemma.~\ref{lemma:equality} has proven that results of ROSC with $W$ prediction window size are the same as the results of applying offline projected gradient descent algorithm with $W$ update times. Hence, we can approximate the optimal dynamic offline algorithm by using ROSC with a large $W=300$.
\end{itemize}

\textbf{Noisy prediction model} Considering predictions are imperfect in practice, we use the the prediction error model in Chen \textit{et al.} \cite{chen2016using} to simulate predictions with noisy errors. In detail, the error at time $\tau$ for the prediction of service $n$ at time $t$ is calculated by $\lambda_{n,t}\sum_{s=\tau}^{t} R e_n(s)$, where $R$ is a noise weight and $e_n(s)$ is per-step noise for service $n$ at time $s$. In the simulations, we let $e_n(s), \forall n,s$ follow standard normal distribution and simulate on various $R$.



\begin{figure*}[t]
    \captionsetup[subfigure]{labelformat=empty,justification=centering,farskip=2pt,captionskip=1pt}
    \centering
    \subfloat[(a) Variable cost ratio]
    {
       \includegraphics[width=0.235\linewidth]{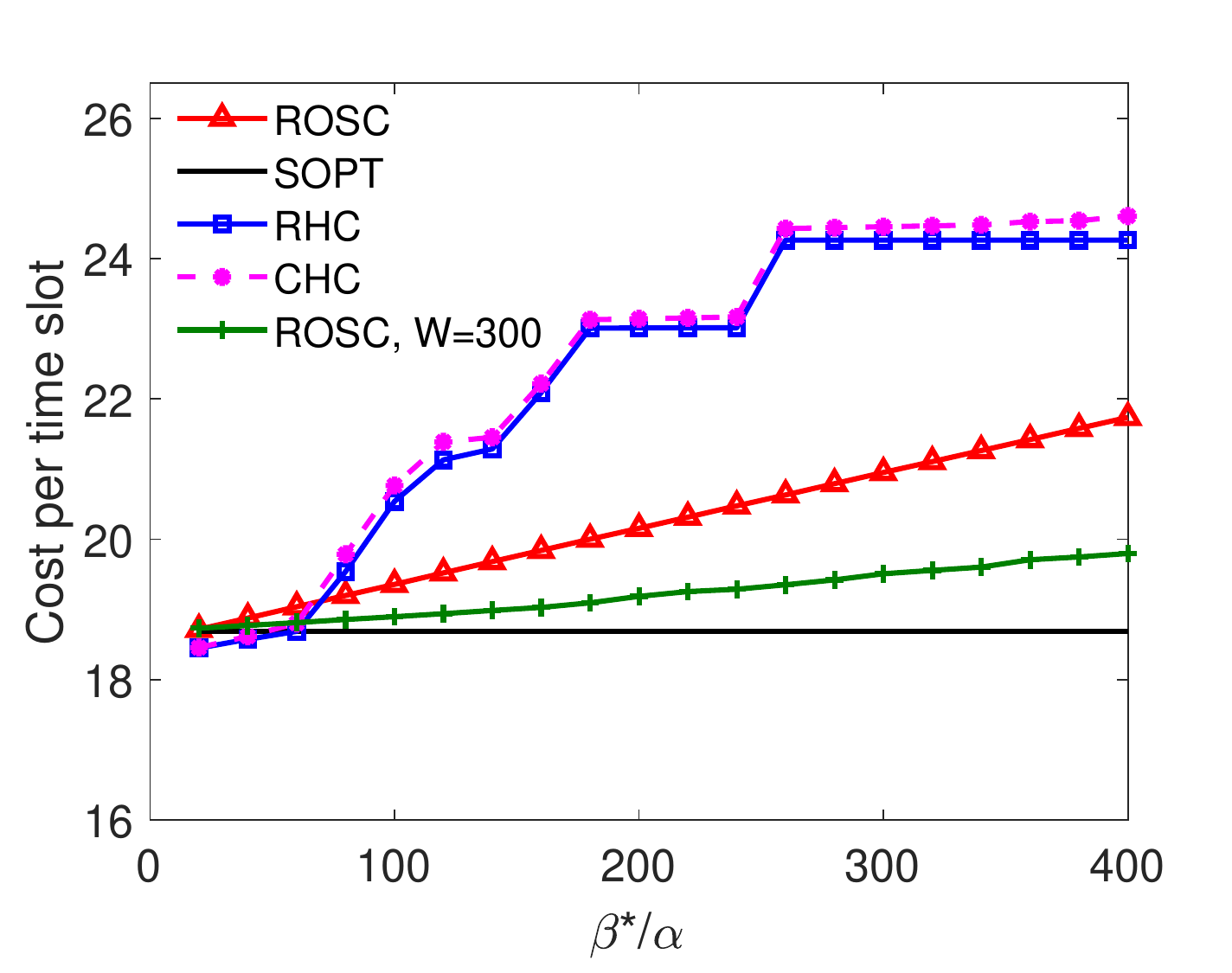}
       \label{fig:replac:beta}
    }
    \subfloat[(b) Variable caching limit]
    {
       \includegraphics[width=0.235\linewidth]{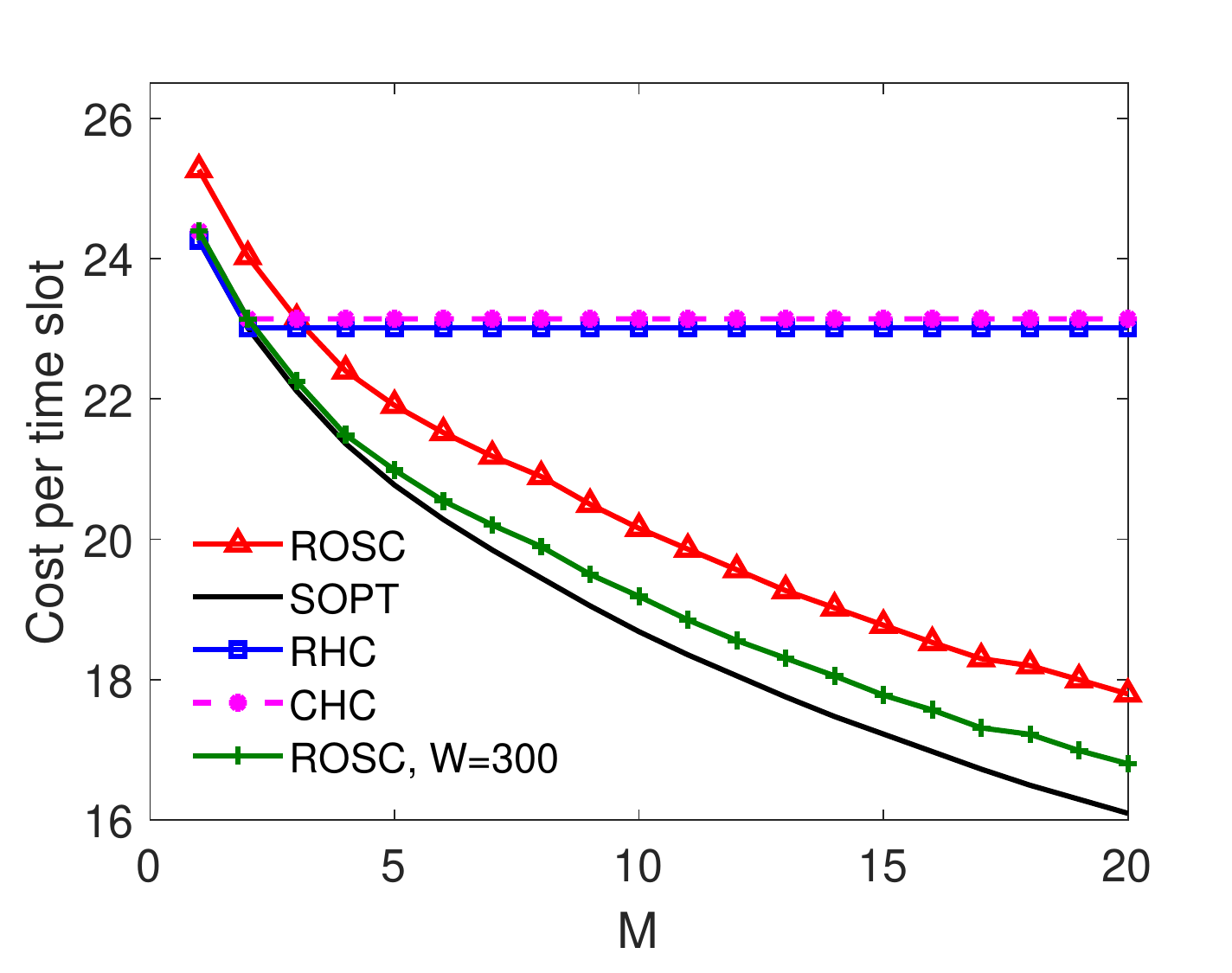}
       \label{fig:replac:m}
    }
    \subfloat[(c) Variable prediction window size]
    {
       \includegraphics[width=0.235\linewidth]{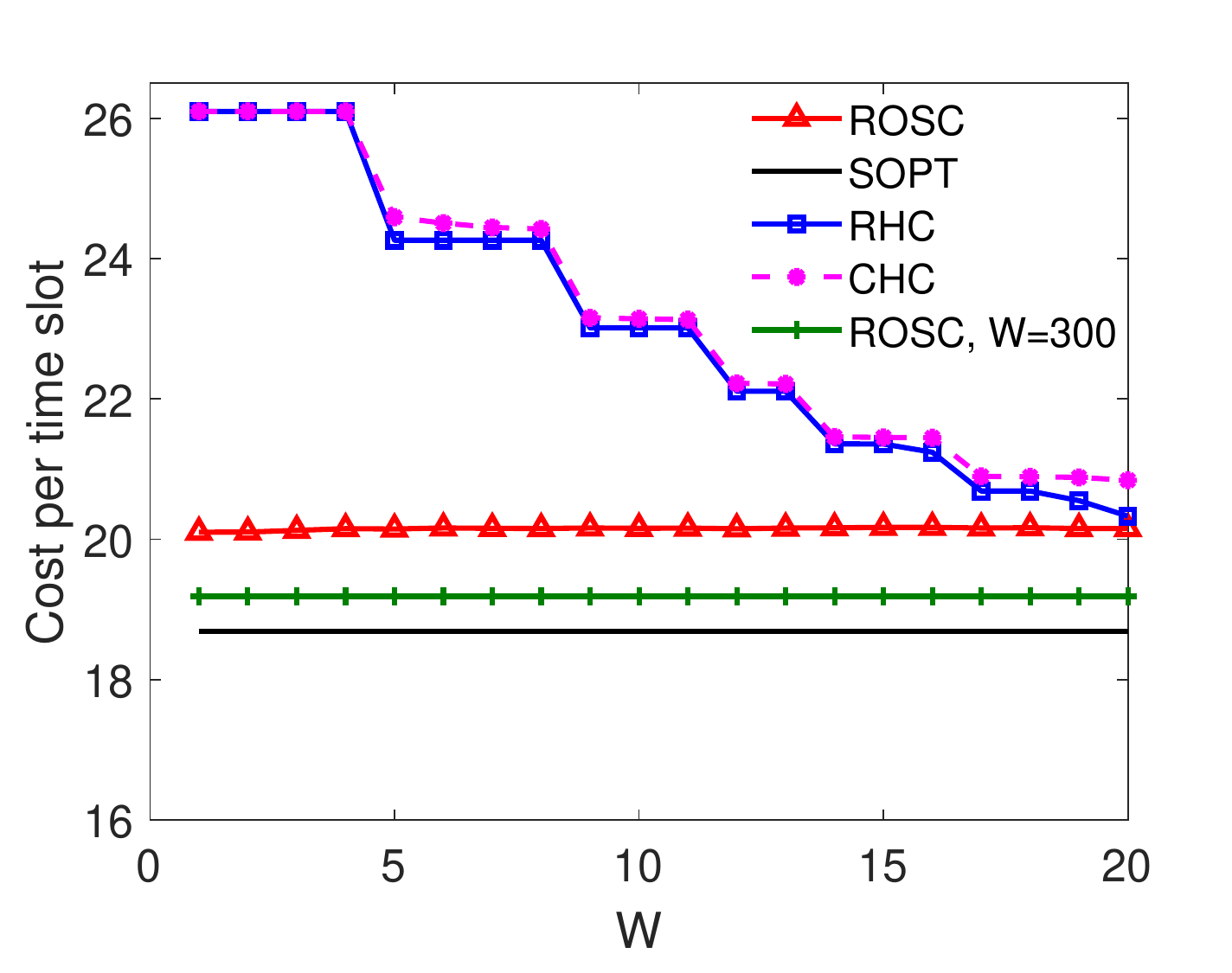}
       \label{fig:replac:w}
    }
    \subfloat[(d) Variable prediction error weight]
    {
       \includegraphics[width=0.235\linewidth]{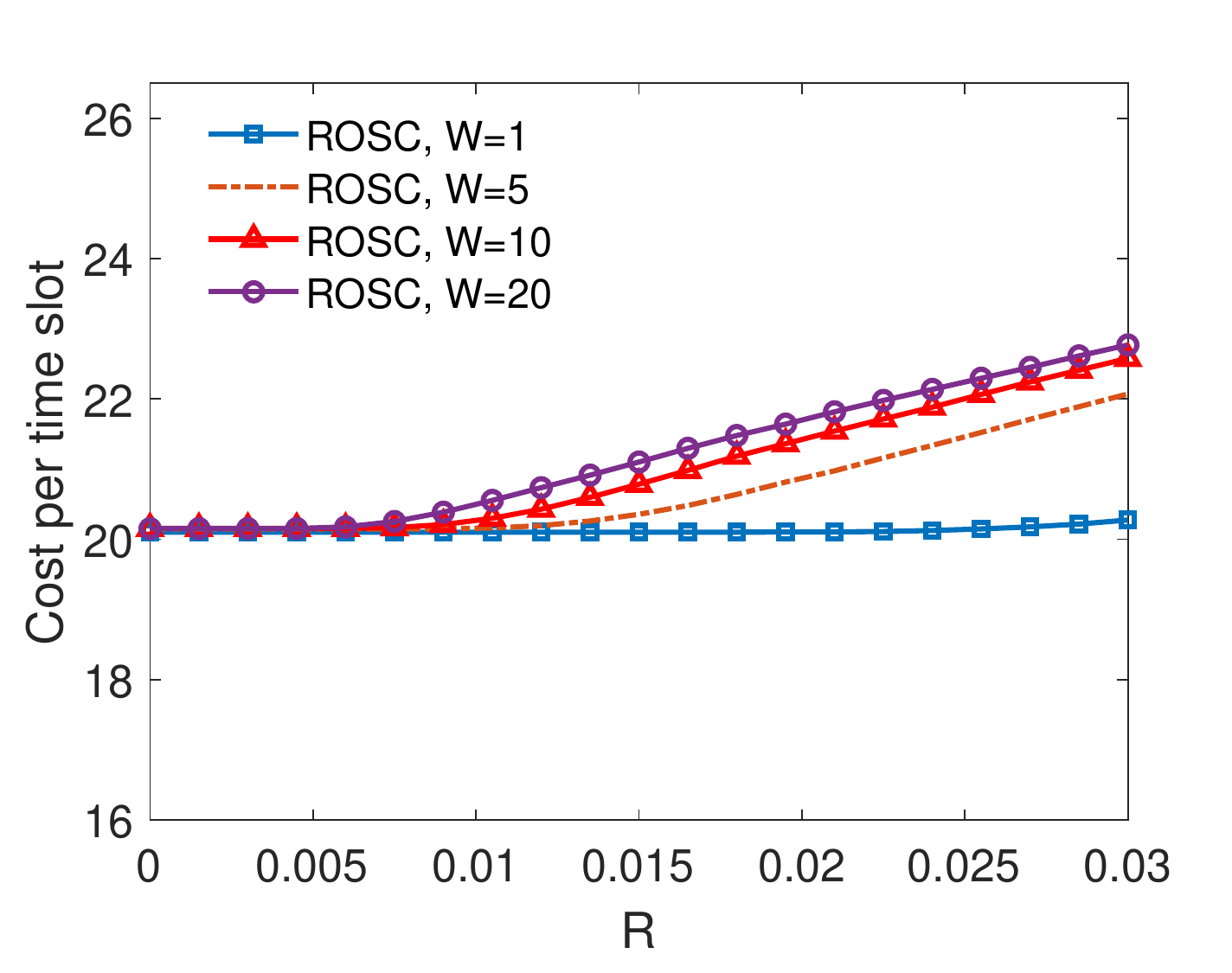}
       \label{fig:replac:r}
    }
    \caption{Simulation results of cost per time slot on the Replacement data set.
    }
    \label{fig:replac}
\end{figure*}

\begin{figure*}[t]
    \captionsetup[subfigure]{labelformat=empty,justification=centering,farskip=2pt,captionskip=1pt}
    \centering
    \subfloat[(a) Variable cost ratio]
    {
       \includegraphics[width=0.235\linewidth]{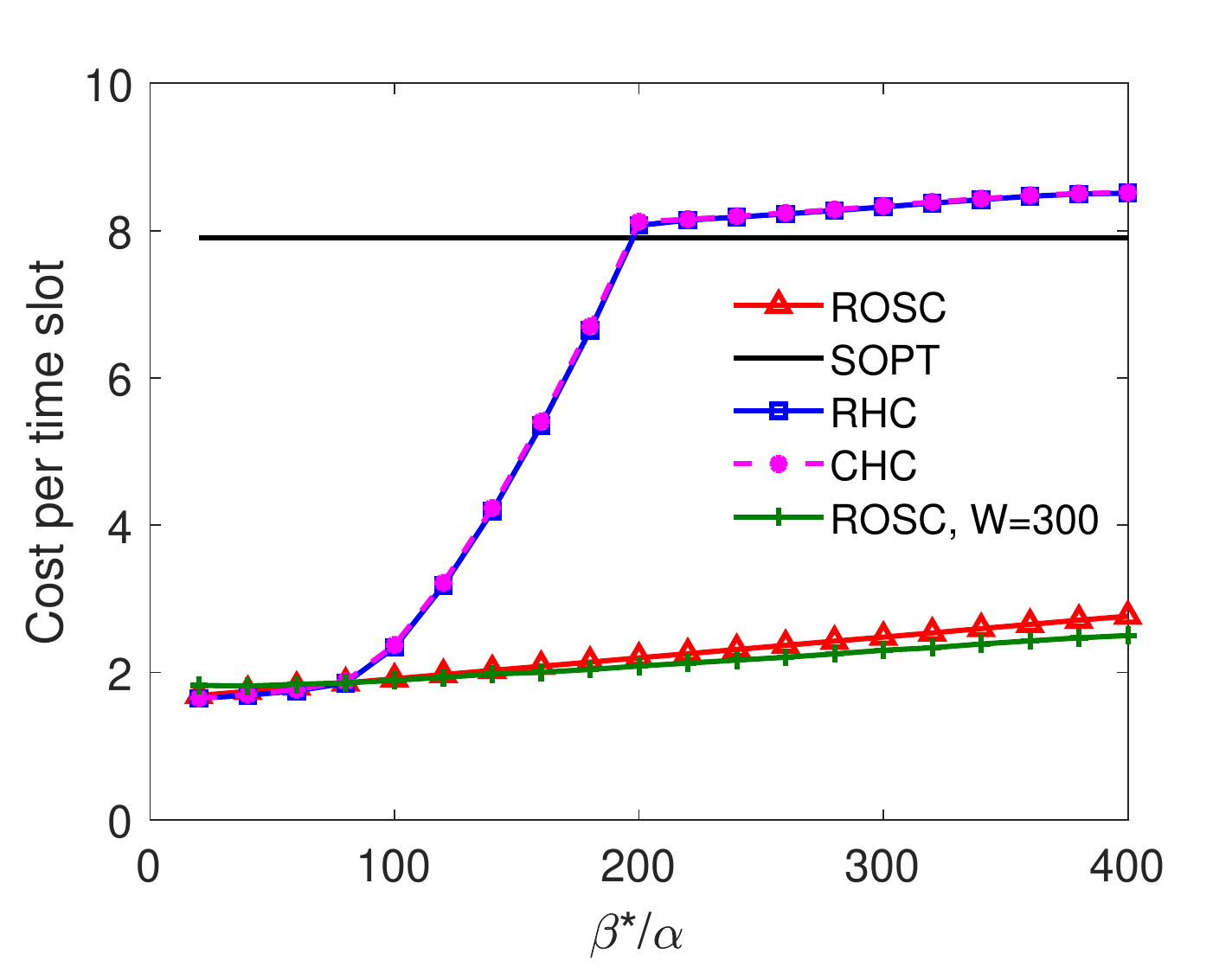}
       \label{fig:poiss:beta}
    }
    \subfloat[(b) Variable caching limit]
    {
       \includegraphics[width=0.235\linewidth]{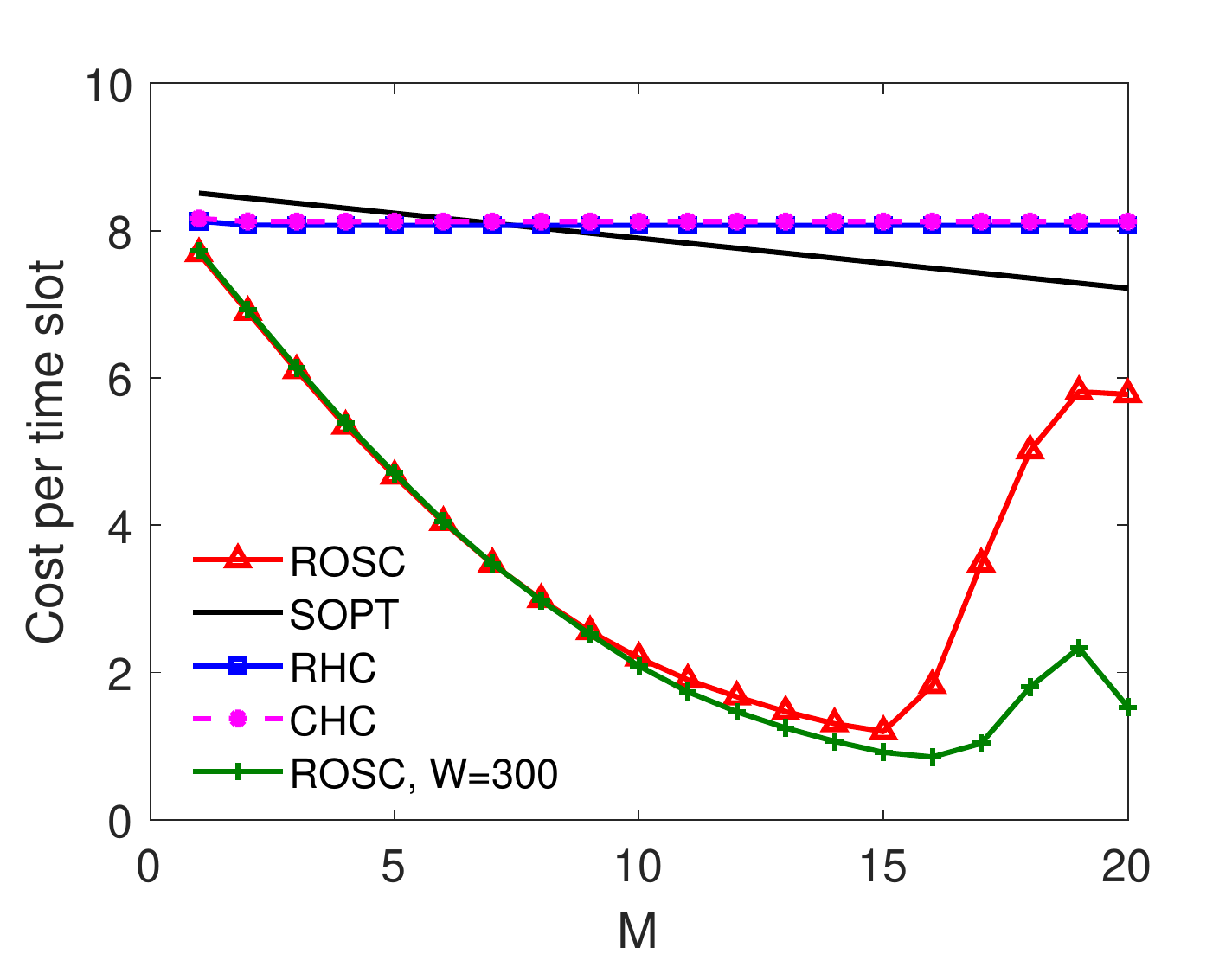}
       \label{fig:poiss:m}
    }
    \subfloat[(c) Variable prediction window size]
    {
       \includegraphics[width=0.235\linewidth]{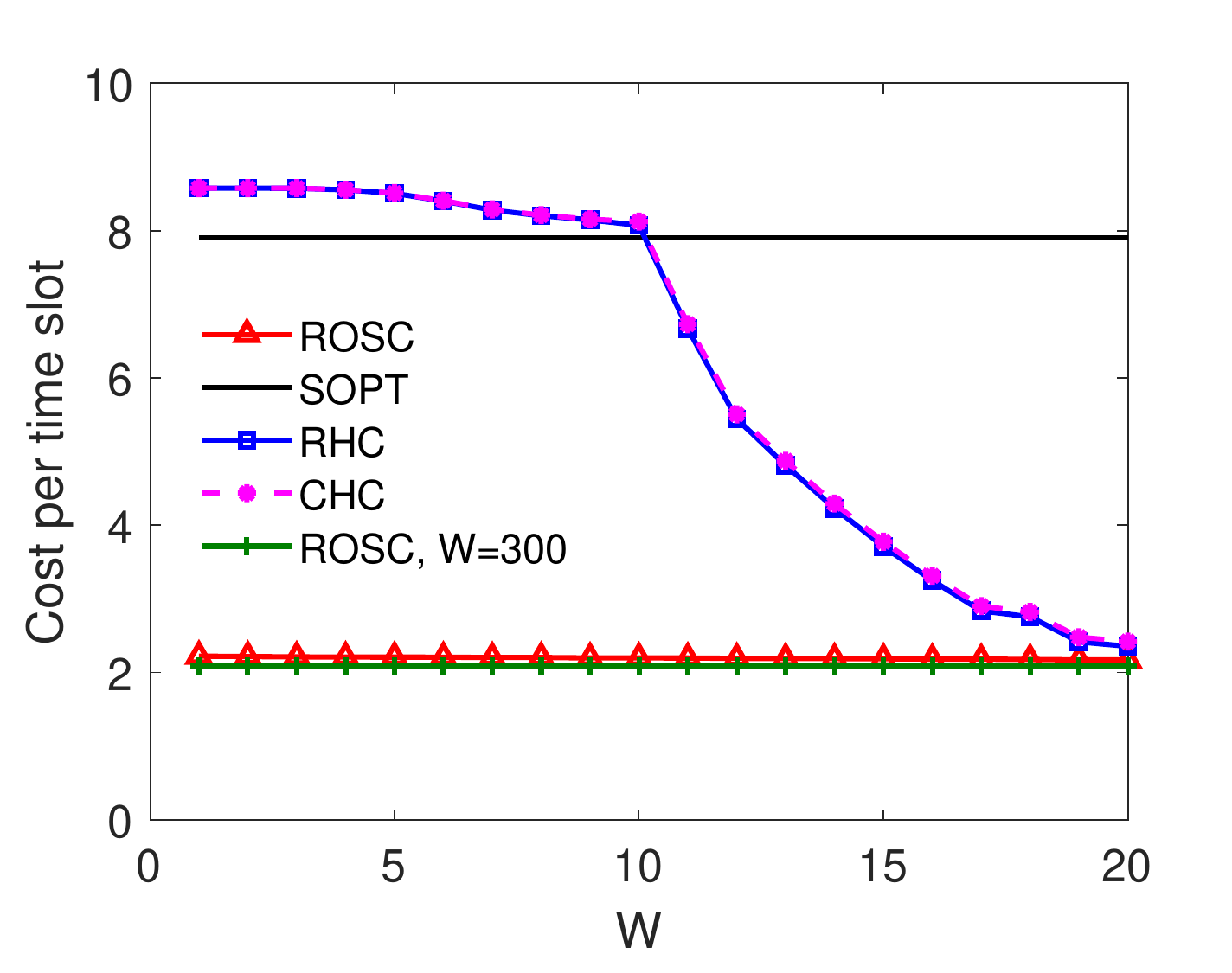}
       \label{fig:poiss:w}
    }
    \subfloat[(d) Variable prediction error weight]
    {
       \includegraphics[width=0.235\linewidth]{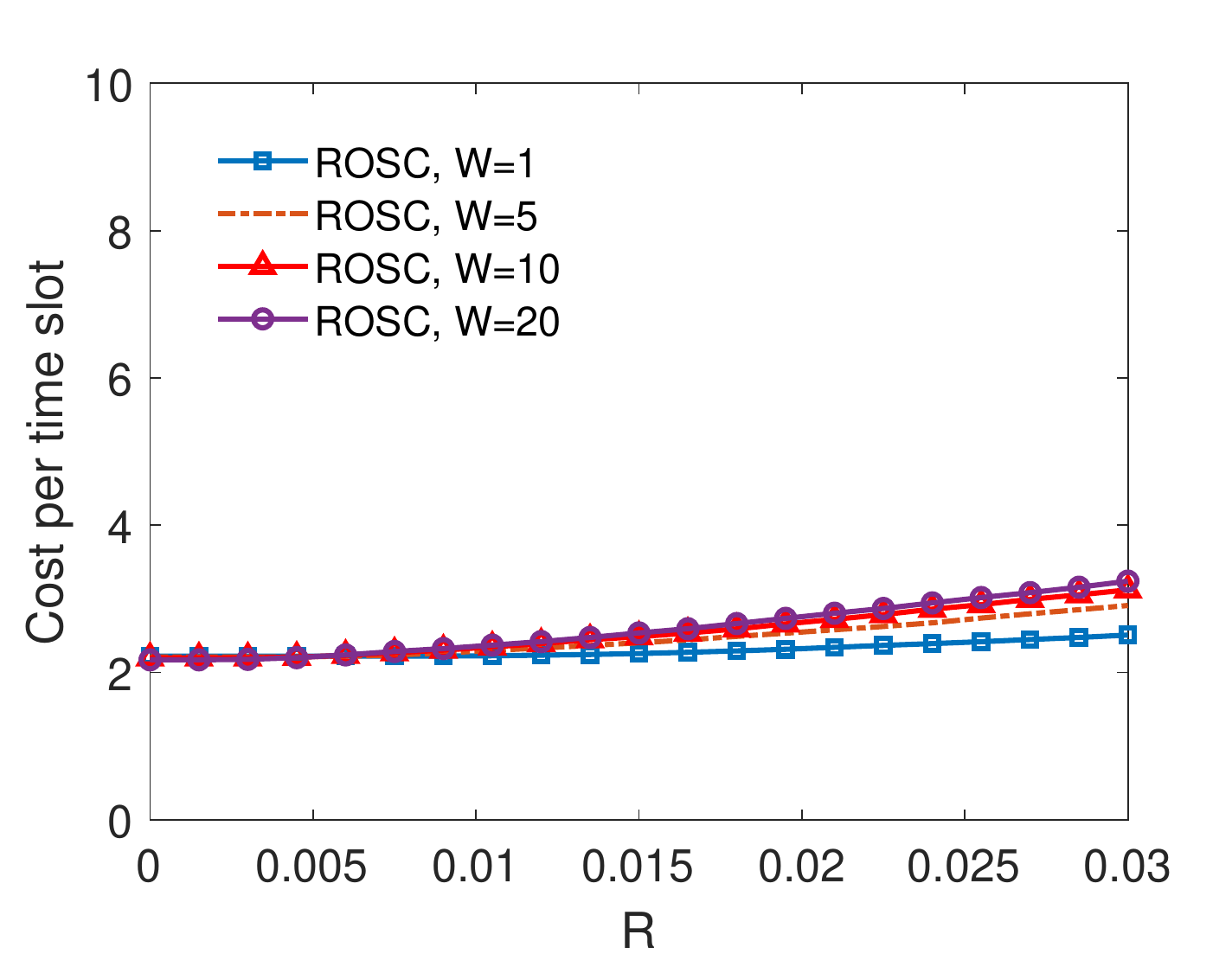}
       \label{fig:poiss:r}
    }
    \caption{Simulation results of cost per time slot on the Poisson data set.
    }
    \label{fig:poiss}
\end{figure*}

\subsection{Evaluation Results}

We present results of our simulations in Table.~\ref{tab2}, Fig.~\ref{fig:replac} and Fig.~\ref{fig:poiss}. Throughout the simulations, parameters are set as $\frac{\beta^*}{\alpha}=200$, $M=10$, $W=10$ and $R=0$ if they are not specified. 
We run 10 independent simulations for each setting and report the average.

Table.~\ref{tab2} evaluates the runtimes of algorithms. It can be seen that ROSC runs much faster than RHC and CHC, and it is less influenced by the increment of the prediction window size $W$. Both RHC and CHC require solving a complex finite-horizon optimization problem with size $O(NW)$, which is why their runtimes increase nearly exponentially as $W$ increases. In contrast, under our ROSC, the runtime is linear in $W$.

\begin{table}[htbp]
\caption{Average runtime of algorithms}
\begin{center}
\begin{tabular}{|c|c|c|c|c|c|c|}
\hline
Algorithm & $W=1$ & 5 & 10 & 15 & 20 \\
\hline
RHC& 426$^{\mathrm{*}}$ & 739 & 1499 & 2585 & 4036\\
\hline
CHC& 855  & 1463 & 2979 & 5100 & 8072\\
\hline
ROSC & 124 & 130 & 137 & 144 & 150\\
\hline
\multicolumn{5}{l}{$^{\mathrm{*}}$ Results are measured in seconds.}
\end{tabular}
\label{tab2}
\end{center}
\vspace{-3mm}
\end{table}

Figs. \ref{fig:replac:beta} -- \ref{fig:replac:r} and \ref{fig:poiss:beta} -- \ref{fig:poiss:w} compare the costs incurred under different algorithms over various settings. It can be observed that RHC and CHC both perform much worse than our ROSC in most cases, especially when $W$ is small. 
Based on the algorithm design, RHC and CHC will only change their caches to host a service $n$ at time $t$ if $\sum_{\tau=t}^{t+W-1}\lambda_{n,\tau}>\frac{\beta^*}{\alpha}$. Hence, when $W$ is small, RHC and CHC are not responsive to gradual changes in long-term trends. It can also be observed that ROSC performs better than the static optimal offline algorithm in the Poisson data set, and has a close performance to SOPT in the replacement data set. In the Poisson data set, the popularity of services changes over time, and no service is always popular. The offline algorithm performs worse than ROSC as it cannot catch the changes in popularity.



Finally, Fig.~\ref{fig:replac:r} and Fig.~\ref{fig:poiss:r} show the result of ROSC with different $W$ under different $R$. It should be noticed that the standard deviation of the prediction error at time $t$ is $WR\lambda_{n,t}$, which increases with both $W$ and $R$. Simulation results show that ROSC is very robust against prediction errors. For example, even when $W=10$ and $R=0.03$, under which case the prediction error is 30$\%$ of the arrival rate, ROSC still outperforms RHC and CHC without prediction error in both data sets. 



\section{Conclusion}
\label{sec:conclusion}

This paper studies an online service caching problem with predictions and analyzes the performance of the proposed algorithm with expected dynamic regret and complexity. In detail, we introduce an auxiliary cost function and then propose a randomized online algorithm, ROSC. ROSC applies an online projected gradient descent step with respect to the auxiliary cost function and uses a randomized algorithm to obtain integer solutions. We show that the expected dynamic regret of ROSC is bounded by the total time horizon and the path length of the requests, which represents changes in requests over time. We further prove that this bound is sublinear with the length of time horizon when the path length is sublinear and parameters are properly chosen. Simulations with two different data sets have shown that ROSC has much better performance than two state-of-the-art algorithms, RHC and CHC, under various parameter settings.

\section*{Acknowledgment}

This material is based upon work supported in part by NSF under Award Number ECCS-2127721, in part by the U.S. Army Research Laboratory and the U.S. Army Research Office under Grant Number W911NF-22-1-0151, and in part by Office of Naval Research under Contract N00014-21-1-2385.

\newpage

\balance

\bibliographystyle{ieeetr}
\bibliography{reference}

\end{document}